# Unlocking the Potential of Binding Corporate Rules (BCRs) in Health Data Transfers[*]

Marcelo Corrales Compagnucci, Mark Fenwick and Helena Haapio

**Abstract:** This chapter explores the essential role of Binding Corporate Rules (BCRs) in managing and facilitating secure health data transfers within corporate groups under the EU General Data Protection Regulation (GDPR). BCRs are tailored to ensure compliance with the GDPR and similar international data protection laws, presenting a flexible mechanism for transferring sensitive health and genomic data. The chapter situates BCRs within the broader spectrum of GDPR's international data transfer mechanisms, addressing the unique challenges posed by the sensitive nature of health data and the increased adoption of AI technologies.

The European Data Protection Board (EDPB) Recommendations 1/2022 on BCRs, issued following the *Schrems II* decision, are critically analyzed, highlighting their stringent requirements and the need for a balanced approach that prioritizes data protection and an AI governance framework. The chapter outlines the BCR approval process, stressing the importance of streamlining this process to encourage broader adoption. It underscores the necessity of a multidisciplinary approach in developing BCRs, incorporating recently adopted international standards and frameworks, which offer valuable guidance for organizations to build trustworthy AI management systems. They guarantee the ethical development, deployment, and operation of AI, which is essential for its successful integration and the broader digital transformation.

In conclusion, BCRs are positioned as essential tools for secure health data management, fostering transparency, accountability, and collaboration across international borders. The chapter calls for proactive measures to incentivize BCR adoption, streamline approval processes, and promote more innovative approaches, ensuring BCRs remain a robust mechanism for global data protection and compliance.

---

[*] This is a working paper. The final version will be submitted to the forthcoming book *International Transfers of Health Data: A Global Perspective*, Corrales Compagnucci, M. and Fenwick, M. (eds) 2025, Perspective in Law, Business and Innovation series, Springer Nature, Singapore (currently under review).

# 1 Introduction

International data transfers across multiple sectors of economic and social life pose a substantial challenge for global organizations, particularly in the context of stringent data protection regulations such as the European Union (EU) General Data Protection Regulation (GDPR).[1] This challenge is amplified and especially acute when it comes to sensitive health data transfers and the increased adoption of AI technologies, given the complexities and heightened privacy concerns associated with this type of personal information.[2]

AI plays a crucial role in precision medicine by analyzing vast health data sets to create personalized treatments, necessitating secure international health data transfers. For example, it identifies genetic mutations for tailored therapies, predicts treatment responses based on comprehensive data, accelerates drug discovery for rare diseases through global collaboration, enhances medical imaging for optimal treatment paths, and supports real-time monitoring for conditions like diabetes. These applications underscore AI's potential in delivering personalized healthcare while highlighting the need for secure global data sharing to facilitate research and innovation.[3]

Under Chapter 5 of the GDPR, there are several data transfer mechanisms, and it is essential to assess the specifics of each data transfer and choose the most suitable GDPR-compliant method to safeguard personal data during any international transfer of such data.[4] In this chapter, our primary emphasis will be on data transfers relying on Binding Corporate Rules (BCRs) as a mechanism, particularly in the context of genomics and health-related data transfers.

BCRs exhibit a certain degree of flexibility in comparison with other mechanisms. In particular, they can be individually tailored to cover specific data types. When it comes to the transfer of health and genomic data to third countries or international organizations, BCRs have limited applicability within the genomics community. The primary challenge in genomics data sharing, therefore, lies not within organizations but *between* them. This is important because genomics in healthcare or research relies on collaboration among *interdependent* organizations rather than global-spanning corporations. Nevertheless, BCRs remain valuable tools for organizations positioned to leverage them. Moreover, some organizations have explored the potential of establishing subsidiaries of non-European Economic Area (EEA) based organizations within the EEA to capitalize on the flexibility offered by BCRs.[5]

In November 2022, the European Data Protection Board (EDPB) issued preliminary guidance, namely EDPB Recommendations 1/2022, on BCRs for data transfers. These recommendations were in alignment with the Court of Justice of the EU (CJEU) *Schrems II* decision,[6] which is a significant legal ruling related to international data transfers.[7] In *Schrems II*, the CJEU underscored the importance of implementing measures that offset the absence of robust data protection in a third country through the establishment of suitable safeguards for data subjects.[8]

---

[1] Regulation (EU) 2016/679 of the EP and of the Council of 27 April 2016 on the protection of natural persons with regard to the processing of personal data and on the free movement of such data, and repealing Directive 95/46/EC, OJ 2016 L 119, 1 (General Data Protection Regulation, GDPR).
[2] Panagopoulos et al. (2022).
[3] Corrales Compagnucci et al. (2022).
[4] Jurcys, Corrales Compagnucci and Fenwick (2022).
[5] PHG Foundation, The GDPR and Genomic Data: The Impact of the GDPR and DPA 2018 on Genomic Healthcare and Research. A PHG Foundation Report Founded by the Information Commissioner's Office, pp. 119-120.

However, there is room for improvement in offering a more comprehensive and forward-thinking perspective. The guidance introduced stringent requirements that exceeded the demands of the new Standard Contractual Clauses (SCCs), which are another commonly used transfer mechanism for local law assessments and the handling of government access requests. To enhance the effectiveness of BCRs, it is therefore essential to recommend adopting a more flexible approach that emphasizes the importance of balancing data protection and adaptability with new technologies such as AI, and this objective needs to be at the core of the BCRs.

To fully harness the potential of BCRs, it is therefore imperative to encourage their widespread adoption, streamline the approval process, and integrate new standards and guidelines for adopting AI. These measures can foster collaboration, drive favorable outcomes, create opportunities, and proactively and ex-ante address potential issues and concerns.

Following this introduction, this chapter is organized as follows. Section 2 provides a concise overview of the various international data transfer mechanisms governed by the GDPR. Section 3 explores the essential provisions of BCRs and outlines the BCR approval process. Section 4 critically analyzes the EDPB Recommendations 1/2022 concerning BCRs for data transfers. Section 5 discusses recommended international standards and frameworks for managing data security and AI-related risks. Section 6 concludes.

## 2 International Data Transfer Mechanisms under the GDPR

In this section, we will provide an overview of the various data transfer mechanisms available under the GDPR. These mechanisms can be visualized as a three-tiered pyramid, with BCRs occupying the second level, immediately following so-called adequacy decisions. The tools for facilitating the secure international transfer of personal data from the EEA are detailed in Chapter V of the GDPR. These methods include:[9]

1. **Transfers based on an 'adequacy decision' by the European Commission (Art. 45 GDPR):** This allows data to be transferred to countries or regions that the European Commission deems to provide an adequate level of data protection. For instance, if the European Commission determines that Country X has data protection regulations and practices in place that closely align with the GDPR's high standards and principles, organizations in the EU can freely transfer personal data to Country X without the need for any additional safeguards or agreements.[10]

---

[6] C-311/18 Data Protection Commissioner v Facebook Ireland Ltd and Maximillian Schrems [2020] ECLI:EU:C:2020:559 (*Schrems II*).
[7] For a full recount of the *Schrems I* and *Schrems II* cases, see Section 3 in Chapter 2 of this book.
[8] Corrales Compagnucci et al. (2020), pp. 153-160.
[9] Minssen et al. (2020), p. 37.

2. **Transfers subject to 'appropriate safeguards' by the controller/processor, provided that enforceable data subject rights and effective legal remedies are available (Art. 46, Art. 47 GDPR):** This option allows data transfers when appropriate measures are in place to protect data subjects' rights and provide legal recourse if needed. For instance, a research institution in the EEA may need or want to transfer sensitive genetic data to a third-party research organization located outside the EEA for some collaborative studies. To ensure the protection of this genetic data, the two organizations may establish a legally binding contract incorporating robust data protection provisions and mechanisms designed to address any potential breaches. This approach serves to uphold the rights of data subjects and establishes a legal recourse for addressing any violations concerning any includeded sensitive genetic information.[11]

Among the most commonly used transfer tools for ensuring the lawful transfer of personal data under the GDPR are:

**Standard Contractual Clauses (SCCs):** These are standardized, pre-approved contractual agreements provided by the European Commission. SCCs contain specific provisions that outline data protection measures and safeguards.[12] There are also ad-hoc contractual clauses. Unlike the standard SCCs provided by regulatory authorities, ad-hoc clauses are tailored to meet particular needs or scenarios not covered by the standard versions. These ad-hoc clauses must be approved by the data protection authority (DPA).

**Binding Corporate Rules (BCRs):** These are internal data protection rules established by multinational organizations that align with GDPR standards. BCRs are particularly valuable for large corporations with subsidiaries or branches operating across international borders.[13]

---

[10] To date, the European Commission has granted adequacy recognition to the following countries and regions: Andorra, Argentina, Canada (for commercial organizations), Faroe Islands, Guernsey, Israel, Isle of Man, Japan, Jersey, New Zealand, Republic of Korea, Switzerland, the United Kingdom under both the GDPR and the LED, the United States (specifically, commercial organizations participating in the EU-US Data Privacy Framework), and Uruguay. These entities have been deemed to offer a sufficient level of data protection by the European Commission. See, Adequacy Decisions: How the EU determines if a non-EU country has an adequate level of data protection, available at: https://commission.europa.eu/law/law-topic/data-protection/international-dimension-data-protection/adequacy-decisions_es?etrans=it (Accessed 3 January 2024). See also, Report from the Commission to the European Parliament and The Council on the first review of the functioning of the adequacy decisions adopted pursuant to Article 25(6) of Directive 95/46/EC, SWD(2024) 3 final. Brussels, 15.1.2024.
[11] Corrales Compagnucci et al. (2024), pp. 151-172.
[12] Art. 46 GDPR. For a detailed account on Standard Contractual Clauses see, Corrales Compagnucci, Aboy and Minssen (2021).
[13] Art. 47 GDPR. See, e.g., Kuner (2021), pp. 194-198.

**Codes of Conduct:** Organizations can develop industry-specific codes of conduct or adhere to certification mechanisms that align with GDPR requirements. These provide additional assurances for data transfers.[14] For instance, the GDPR provides a framework for harmonizing data protection laws in Europe and allows for the creation of codes of conduct tailored to specific areas like biobanking,[15] genomic and health-related data sharing.[16]

**Certifications:** An approved certification mechanism, coupled with binding and enforceable commitments from the controller or processor in the third country, ensures the application of necessary safeguards, including the protection of data subjects' rights.[17]

3. **Derogations for specific situations (Art. 49 GDPR):** These are exceptions allowing data transfers without specific safeguards in cases like obtaining explicit consent from data subjects.[18] For instance, if one considers a EEA-based medical research organization collaborating with an overseas institution to conduct a vitally important clinical study on a global health crisis. In this case, due to the urgency of the situation and the necessity for quickly sharing such sensitive health data, the derogation could be justified and applied. While explicit consent from all individuals might be impractical, data transfers can proceed under such an exception, as they are deemed vital for the joint research project and have the laudable purpose of addressing the health crisis. Other derogation examples are inter alia when the transfer is necessary for the performance of a contract between the data subject and the data controller (i.e., 'necessity for contract performance'),[19] or when the transfer is necessary for reasons of substantial interest, with a clear legal basis ('substantial public interest'),[20] and when the transfer is necessary to protect the vital interests of the data subject or other individuals ('vital interest').[21]

As such, the GDPR provides a robust and durable framework that offers a number of clear mechanisms for international data transfers, as well as some clearly defined exceptions where derogation from the general requirement is permissible.

---

[14] Art. 40 and 46 (2) (e) GDPR.
[15] Krekora-Zając, Marciniak and Pawlikowski (2021).
[16] Phillips et al. (2020).
[17] Art. 42 and 46 (2) (f) GDPR. Examples of certifications are HITRUST Framework, ISO 27001 and ISO 27701.
[18] Art. 49 (1) (a) GDPR. See, e.g., Fedeli et al. (2022).
[19] Art. 49 (1) (b) GDPR.
[20] Art. 49 (1) (d) GDPR.
[21] Art. 49 (1) (f) GDPR.

# 3 Key Provisions of BCRs and the Approval Procedure

In the following, we focus more specifically on BCRs as a mechanism. BCRs are an innovative and comprehensive data transfer mechanism developed collaboratively by the European Commission, member state data protection authorities, and multinational businesses.[22] They are essentially data protection policies adopted by EEA companies for transferring personal data outside the EEA within their group of undertakings or enterprises[23] when they are engaged in a joint economic activity.[24] As such, BCRs facilitate responsible cross-border data flows within a corporate group, serving as a so-called 'gold standard' for privacy and data protection management programs within a shared organizational group.[25]

Organizations, whether acting as controllers or processors,[26] that have implemented BCRs report various business benefits and gain internal and external recognition for their 'BCR status.' This recognition acts as a 'soft certification,' showcasing an organization's commitment to and compliance with data privacy regulations to both business partners and individuals. Furthermore, BCRs are recognized as a valid international data transfer mechanism by several other non-EEA member countries, including the U.K., Singapore, Brazil, and South Africa. [27]

These rules encompass all general data protection principles and enforceable rights to ensure adequate safeguards for data transfers. They are legally binding and must be upheld by *every* member of the group involved.[28] As such, they serve as a fundamental baseline framework for multinational organizations involved in international data transfers, ensuring compliance with data protection regulations and safeguarding individuals' personal data. To achieve this goal, BCRs must meet the following specific criteria:[29]

1. **Legally Binding and Universal Applicability**: BCRs are legally binding (both internally and externally) and universally applicable, extending to every member concerned within a group of undertakings or enterprises engaged in a joint economic activity, and this also includes employees with access to personal data.

---

[22] Bellamy (2023).

[23] An enterprise refers to 'a natural or legal person engaged in an economic activity, irrespective of its legal form, including partnerships or associations regularly engaged in an economic activity' (Article 4(18) GDPR).

[24] Recital 110 GDPR.

[25] Pryke (2022).

[26] A 'controller' is defined as the entity responsible for 'determining the methods and objectives of data processing' (as per Article 4(7) of the GDPR). Meanwhile, a 'processor' refers to the entity handling personal data on behalf of the controller, particularly when the controller has outsourced the data processing task (as per Article 4(8) of the GDPR). For a more detailed account and understanding of these roles, see Dahi and Corrales Compagnucci (2022).

[27] Bellamy (2023).

[28] Binding Corporate Rules (BCRs): Corporate Rules for Data Transfers within Multinational Companies, available at: https://commission.europa.eu/law/law-topic/data-protection/international-dimension-data-protection/binding-corporate-rules-bcr_en (Accessed 8 July 2024).

[29] Art. 47 GDPR.

2. **Empowering Data Subjects**: BCRs expressly confer enforceable rights on data subjects regarding the processing of their personal data, empowering individuals to assert full control over their data if they so choose.
3. **Detailed Specifications**: BCRs provide comprehensive specifications, including:
    - *Organizational Structure*: Clearly defining the structure and contact details of the group of undertakings or enterprises engaged in a joint economic activity and its individual members.
    - *Data Transfer Specifics*: Detailing data transfers, including categories of personal data (such as health-related sensitive data), processing types and purposes, affected data subject categories, and identification of relevant third countries.
    - *Data Protection Principles*: Ensuring adherence to general data protection principles such as purpose limitation, data minimization, limited storage periods transparency, data quality, data protection by design and by default, the legal basis for data processing, processing of special categories of personal data, measures to ensure security, and the requirements concerning onward transfers to entities not subject to the BCRs.
4. **Monitoring and Compliance**: BCRs establish mechanisms for monitoring compliance, incorporating elements like data protection audits, training protocols, and complaint-handling mechanisms. These measures facilitate corrective actions to protect data subject rights and ensure overall compliance.
5. **Cooperation with Supervisory Authorities**: BCRs outline cooperation mechanisms with supervisory authorities, ensuring collaboration for compliance within the organization.
6. **Training and Awareness**: BCRs incorporate provisions for appropriate data protection training, targeting personnel with access to personal data. This helps raise awareness and ensures that individuals handling data are well-informed about data protection principles.

By adhering to these requirements, organizations can not only ensure legal compliance but also promote transparency, accountability, and data subject rights within the context of international data transfers. BCRs serve as a vital tool in facilitating responsible and ethical data handling on a global scale.

More specifically, there are two primary types of BCRs:[30]

1. Controller BCRs (BCR-C): These are suitable for data transfers between controllers within the EEA and other group company controllers or processors located outside the EEA. They apply to entities within the same group acting as controllers, as well as those functioning as 'internal' processors.

---

[30] PWC, Binding Corporate Rules, available at: https://www.pwc.com/m1/en/publications/documents/pwc-binding-corporate-rules-gdpr.pdf (Accessed 8 July 2024).

2. Processor BCRs (BCR-P): Processor BCRs are relevant when personal data is received from an EEA-based controller that is not part of the group. Group members then process this data as processors or sub-processors. These BCRs serve as an alternative to incorporating the SCCs into service agreements with controllers.

The European Commission is empowered to specify the format and procedures for exchanging information among controllers, processors, and supervisory authorities concerning BCRs. Moreover, companies seeking approval for their BCRs must submit them to the relevant EU data protection authority for vetting. The authority will assess and approve the BCRs following the process outlined in Article 63 of the GDPR.[31] This procedure may involve multiple supervisory authorities if the group applying for BCR approval operates in more than one EU Member State. Once finalized in line with the EDPB's opinion, the competent authority will then grant approval for the BCRs.[32]

While BCR approval ensures lawful personal data transfers to third countries, it is essential to note that BCRs can also function as a global data protection policy for multinational groups. Nevertheless, the scope of BCR approval remains confined to transfers from GDPR-bound entities to third countries and their onward transfers to other BCR members outside the EEA.

As such, BCR approval does not entail an assessment of the GDPR compliance of each processing activity. Data exporters must individually ensure adherence to GDPR requirements, such as the lawfulness of processing and compliance with Article 28 (pertaining to transfers to processors). Additionally, data exporters must evaluate the need for supplementary measures on a case-by-case basis to guarantee a level of data protection essentially equivalent to that provided by the GDPR.[33]

Understanding the BCR approval process and the scope of BCRs is now a crucial element for organizations involved in international data transfers. This knowledge ensures compliance with data protection regulations and facilitates secure and efficient exchange of personal data across borders.

However, it is also important to note that BCR approval is typically a lengthy and resource-intensive endeavor and process. BCR implementations can create a significant administrative burden and prolong the regulatory application and approval process, which will often leave organizations in a legal limbo during the protracted waiting period.

Due to these challenges, many organizations hesitate to adopt BCRs and instead opt for potentially less effective contracts for their internal data transfers. As a result, BCRs have not yet fully realized their potential.[34] For example, the UK's Information Commissioner's Office (ICO) points out that even a seemingly straightforward application can take up to twelve months before receiving approval. Therefore, organizations need to be of substantial size and already engaged in economically valuable data transfers to justify pursuing BCRs.[35]

---

[31] See also recital 135 GDPR "Consistency Mechanism".

[32] Binding Corporate Rules (BCRs): Corporate Rules for Data Transfers within Multinational Companies, available at: https://commission.europa.eu/law/law-topic/data-protection/international-dimension-data-protection/binding-corporate-rules-bcr_en (Accessed 8 July 2024). A list of BCRs approved under the GDPR is available here: https://edpb.europa.eu/our-work-tools/accountability-tools/bcr_en (Accessed 8 January 2024).

[33] Recommendations 1/2022 on the Application for Approval and on the elements and principles to be found in Controller Binding Corporate Rules (Art. 47 GDPR).

[34] Bellamy (2023).

## 4 EDPB Recommendation

In its December 2022 Recommendation, the EDPB provided draft guidance with updated interpretations and requirements for the BCRs transfer mechanism. Regrettably, however, the EDPB missed an important and valuable opportunity to comprehensively address BCRs with a strategic and forward-thinking approach, preventing the enhancement of this essential transfer mechanism into a more scalable and globally applicable tool for sustainable international data transfers.

In light of the GDPR and recent legal developments, along with the evolving landscape of international data transfers in Europe and beyond, it is imperative to reevaluate and improve BCRs. To effectively harness the potential of BCRs, policymakers should consider the following measure and steps:[36]

1. **Advocate for, incentivize, and recognize their distinctive characteristics.** It is essential for the European Commission, the EDPB, and other supervisory authorities to take proactive measures to encourage more widespread BCR adoption, streamlining the process and making it more appealing for corporate groups of all sizes to attain BCR approval. A BCR differs significantly from a conventional contract, in the sense that it more closely resembles an enforceable corporate code of conduct or a third-party certified accountability framework. As such, it serves as clear evidence of accountability and the presence of a comprehensive data privacy management program across the whole corporate group and all its related entities.[37]
2. **Streamline and transform the approval process.** The BCR requirements should not impose stricter conditions than those governing other transfer mechanisms, such as the SCCs.[38] Instead, they should be tailored to the unique nature of this mechanism. BCRs should be designed to accommodate organizations of all sizes and corporate structures, ensuring scalability and configurability. In this respect, Data Protection Authorities (DPAs) play a crucial role in this process by minimizing administrative burdens and timelines associated with BCR applications while establishing transparent and practical criteria. This approach fosters broader adoption by making BCRs adaptable to the diverse needs of organizations.[39]
3. **Promote a consistent risk-based approach to risk assessments.** BCRs embody a firm dedication to maintaining a uniform standard of privacy protection throughout an entire corporate group. It is imperative for policymakers to align the guidance and criteria for transfer risk assessments under BCRs with those of other transfer mechanisms, such as SCCs, to avoid imposing disproportionately higher standards. Consistency in applying a risk-based, contextual approach across all transfer mechanisms, including BCRs, is therefore essential. Failure to do so would seriously penalize and discourage businesses that have committed to higher compliance standards through BCRs, undermining future incentives for such investments.[40]

---

[35] PHG Foundation, The GDPR and Genomic Data: The Impact of the GDPR and DPA 2018 on Genomic Healthcare and Research. A PHG Foundation Report Founded by the Information Commissioner's Office, p. 119.
[36] Bellamy (2023).
[37] Bellamy (2023).
[38] See, e.g., Corrales Compagnucci, Aboy and Minssen (2021), pp. 37-47.
[39] Bellamy (2023).

4. **Recognize and validate inter-BCR transfers among approved entities.** Currently, corporate groups equipped with BCRs can only utilize them for intragroup data transfers, restricted to controllers and processors within the group. However, considering that BCRs undergo rigorous review and approval by regulators under the GDPR, serving as a comprehensive compliance framework, regulators should enable companies with BCRs to transfer data to other BCR-approved organizations. This recognition would significantly enhance the attractiveness of BCRs. Enabling one BCR-approved organization to exchange data with another BCR-approved entity would mutually benefit both, thus ensuring a shared commitment to privacy and offering substantial reassurance.[41]

## 5 Recommended International Standards and Frameworks to Manage Data Security and AI-related Risks

BCRs are a vital management tool for addressing significant legal challenges related to data transfers.[42] Traditionally, BCRs have focused on legal compliance and risk mitigation. However, it is crucial to recognize that BCRs must be continually updated to address emerging risks, especially with the advent of new technologies such as AI.

AI systems present several risks, particularly concerning trust and privacy across various domains. One significant threat is 'training data poisoning,' where adversaries insert malicious data into training sets to manipulate AI models. This can result in model degradation, bias, security vulnerabilities, and privacy violations. To mitigate these risks, it is essential to implement measures such as data quality assurance, adversarial training,[43] transparency, accountability, and regulatory compliance.[44]

Another significant risk is 'data leakage' in federated learning, where sensitive information from decentralized devices can unintentionally leak due to compromised security protocols, insufficient encryption, or data transmission vulnerabilities. Addressing data leakage requires robust security measures, including advanced encryption techniques, differential privacy mechanisms, and stringent access controls to protect data privacy and maintain the integrity of AI systems.[45]

---

[40] Bellamy (2023).
[41] Bellamy (2023).
[42] Chapter 4 GDPR.
[43] Adversarial training is a technique used to enhance the robustness of AI models against adversarial attacks. It involves exposing the model to adversarial examples—deliberately modified inputs designed to deceive the model—during the training phase. By learning from these adversarial examples, the model becomes better equipped to recognize and resist similar attacks in real-world scenarios. This process improves the model's security and reliability, ensuring it performs accurately even when faced with malicious attempts to compromise its functionality. See, Chen and Hsieh (2023).
[44] See, OWASP, available at: https://genai.owasp.org/llmrisk/llm03-training-data-poisoning/ (Accessed 21 January 2024).

The development and implementation of BCRs require a multidisciplinary approach extending beyond the realm of traditional legal practice. While regulators undoubtedly play a vital role in ensuring BCRs adhere to the necessary GDPR standards and boundaries, the complex nature of BCRs transcends the legal domain. In this way, international standards and frameworks can improve the technical requirements needed to implement them effectively.

By incorporating these standards and frameworks into the BCRs, organizations can harness the power of AI technologies, especially in the field of health data transfers, where complexities and nuances abound. Moreover, BCRs inherently involve various facets of an organization, including governance structure, risk management procedures, and employee training protocols. These standards and frameworks can streamline complex interdependencies, making it easier for individuals without a legal background to understand and engage with the process.

To effectively manage and mitigate the risks associated with AI technologies in international health data transfers, several standards and frameworks are recommended as follows:[46]

1. **ISO/IEC 42001:2023 – Artificial Intelligence Management System**.[47] This comprehensive framework guides organizations on the responsible and effective use of AI. It emphasizes governance through an AI Management System (AIMS), covering ethical considerations, transparency, and continuous learning. Utilizing the Plan-Do-Check-Act methodology, it sets policies, objectives, and processes for responsible AI management and ensures compliance with legal and regulatory standards.
2. **ISO 31000:2018 Risk Management – Guidelines**.[48] This standard offers internationally recognized principles for effective risk management across various sectors. It supports a shared understanding of risks, integrates risk management into strategic decision-making, and enhances operational excellence. Adopting ISO 31000 helps organizations proactively manage risks, build stakeholder confidence, and improve performance through a structured risk management framework.
3. **NIST AI Risk Management Framework (NIST AI RMF)**.[49] Developed by the U.S. National Institute of Standards and Technology, this framework provides a structured approach to managing AI risks. The April 2024 draft publication of the AI RMF Generative AI Profile addresses risks specific to generative AI, offering tailored actions and guidance based on input from over 2,500 experts. The guidance highlights 12 key risks[50] and suggests over 400 actions for developers to mitigate these risks effectively.

---

[45] Jin et al. (2021).
[46] Corrales Compagnucci (2024).
[47] ISO/IEC 42001:2023 - Artificial Intelligence Management System (AIMS), available at: https://www.iso.org/standard/81230.html (Accessed 1 July 2024).
[48] ISO/IEC 31000:2018 - Risk Management Guidelines, available at: https://www.iso.org/standard/65694.html (Accessed 1 July 2024).
[49] NIST AI Risk Management Framework (NIST AI RMF), available at: https://www.nist.gov/itl/ai-risk-management-framework (Accessed 1 July 2024).

4. **IEEE 7000-21 Standard Model Process for Addressing Ethical Concerns during System Design (IEEE Std 7000)**.[51] This standard guides the integration of ethical values into system development, covering aspects like transparency, privacy, and fairness. It helps align innovation management processes with ethical considerations and emphasizes stakeholder communication and traceability of ethical values.
5. **ISO/IEC Guide 51 – Safety Aspects: Guidelines for Their Inclusion in Standards**.[52] This guideline supports standard drafters in integrating safety considerations into various technologies and products. It focuses on consistent risk mitigation throughout the product lifecycle, aligning closely with the EU AI Act's risk identification principles.

Incorporating these standards and frameworks into BCRs is crucial for managing AI-related risks effectively. This involves documenting all AI applications and implementing a risk classification framework, which enhances risk management and ensures adherence to ethical and regulatory requirements. Given the complexity of advanced technologies, a tailored approach is necessary for each application, integrating multiple standards to address specific needs.

When integrating standards and frameworks into BCRs, it is essential to reassess the existing risk management framework to effectively incorporate new AI-related risk strategies. This includes evaluating whether AI implementation may exacerbate existing risks or introduce new ones, necessitating adjustments to current programs. A tailored risk management framework for AI-related risks is crucial for managing international health data transfers. Additionally, BCRs should incorporate an AI governance framework to assess risk intersections, ensuring cohesive and efficient data transfer programs while mitigating potential harms. This approach enhances the interoperability of AI risk management with operational strategies, addressing security and privacy concerns.

# 6 Conclusion

BCRs play a crucial role in efficiently managing personal data and ensuring the secure transfer of data within corporate entities. They are built upon a robust framework that encompasses governance, risk management, training, and audit standards, forming a resilient compliance infrastructure. However, the advent of new technologies, such as AI, has exacerbated privacy and data security risks. Recognizing the unique challenges posed by these technologies, streamlining and accelerating the approval process, and promoting a consistent risk-based approach are paramount.

---

[50] These 12 risks are: 1) CBRN information, 2) confabulation, 3) dangerous or violent recommendations, 4) data privacy, 5) environmental, 6) human-AI configuration, 7) information integrity, 8) information security, 9) intellectual property, 10) obscene, degrading, and/or abusive content, 11) toxicity, bias, and homogenization, 12) value chain and component integration.
[51] IEEE 7000:2021 - IEEE standard model process for addressing ethical concerns during system design, .available at: https://www.en-standard.eu/ieee-7000-2021-ieee-standard-model-process-for-addressing-ethical-concerns-during-system-design/ (Accessed 1 July 2024).
[52] ISO/IEC Guide 51:2014 – Safety aspects – Guidelines for their inclusion in standards, available at: https://www.iso.org/standard/53940.html (Accessed 1 July 2024).

Incorporating new international standards and frameworks into BCRs, particularly in the context of health data transfers, is an effective strategy that enhances their efficacy and value. While lawyers provide expertise in ensuring legal compliance, a collaborative approach that integrates nuanced technical requirements can fully optimize the creation and management of BCRs. This ultimately benefits the organization as a whole and end-users/patients. Such an approach fosters broader adoption by making BCRs adaptable to the diverse needs of organizations on an international scale.

**Acknowledgement:** Marcelo Corrales Compagnucci received funds from the Inter-CeBIL program (grant agreement number NNF23SA0087056). The opinions expressed are the authors' own and not of their respective affiliations. The authors declare no conflicts of interests.